\documentclass[12pt]{article}
\usepackage{amsmath,amssymb} 
\usepackage{epsfig}

\newcommand{\be}{\begin{equation}}
\newcommand{\ee}{\end{equation}}
\newcommand{\bea}{\begin{eqnarray}}
\newcommand{\eea}{\end{eqnarray}}
\newcommand{\ba}{\begin{eqnarray*}}
\newcommand{\ea}{\end{eqnarray*}}

\newcommand{\Fe}[2]{\Fs{#1}{#2}{z}}
\newcommand{\Fiz}[2]{\Fs{#1}{#2}{\frac{1}{z}}}

\newcommand{\Fh}[2]{\,{}_#1F_#2}
\newcommand{\Fs}[3]{\!\!\left[\begin{array}{c}#1\,;\\#2\,;\end{array}#3\right]}

\newcommand{\FpR}[2]{\Fs{#1}{#2}{\frac{p_{12}-p_{13}}{m^2}}}

\newcommand{\Fup}[2]{\Fs{#1}{#2}{\frac{p^2}{m^2}}}

\begin{document}

%%%%%%%%%%%%%%%%%%%%%%%%%%%%%%%%%%%%%%%%%%%%

\title{
\vskip-2cm{\baselineskip14pt
\centerline{\normalsize DESY~08--150 \hfill ISSN 0418-9833}
%\centerline{\normalsize arXiv:xxxx.yyyy\hfill}
\centerline{\normalsize September 2008\hfill}
}
\vskip2.5cm
{
New relationships between Feynman integrals}
\\}

\medskip

\author{
{\sc O.~V.~Tarasov}\thanks{On leave of absence from
Joint Institute for Nuclear Research,
141980 Dubna (Moscow Region), Russia.}
\\
\\
{\normalsize II. Institut f\"ur Theoretische Physik, Universit\"at Hamburg,}\\
{\normalsize Luruper Chaussee 149, 22761 Hamburg, Germany}
}

\date{}

\maketitle

%%%%%%%%%%%%%%%%%%%%%%%%%%%%%%%%%%%%%%%%%%%%%%%%%%%%%%%%%%%%%%%%%%%%%%%%%%%%%%%

\begin{abstract}
New types of relationships between Feynman integrals are
presented. It is shown that Feynman integrals satisfy functional 
equations connecting integrals with different values of scalar 
invariants  and masses. A  method is proposed for obtaining such 
relations. The derivation of functional equations for one-loop 
propagator- and  vertex - type integrals  is given. It is shown that 
a propagator - type integral can be  written as  a sum of two 
integrals with modified scalar invariants and  one  propagator massless. 
The vertex - type integral  can be written as a sum over vertex  
integrals  with  all but one propagator massless and one external 
momentum squared equal to zero. It is demonstrated  that the functional 
equations can be used for the analytic continuation of Feynman integrals 
to different kinematic domains. 

\medskip

\medskip


\end{abstract}

\newpage
 
\section{A  method for deriving functional equations}
Feynman integrals play an important role in making precise
perturbative predictions in quantum field theory.
As is well-known,  these integrals satisfy recurrence 
relations \cite{Petersson}-
%, \cite{'t Hooft:1972fi}, 
%\cite{Tkachov:1981wb}, 
%\cite{Chetyrkin:1981qh}, 
\cite{Tarasov:1996br}.
In general such relations connect several integrals  
$I_{1,n},...,I_{N,n}$  with   $n$ internal lines and
integrals with a lesser number  of internal lines. 
They may be written in the following form
\begin{equation}
\sum_{i=1}^N Q_i(\{m_j \},\{s_{q}\},\nu_{l},d)
 ~I_{i,~n} = \sum_{\substack{ r<n  \\ k}} R_{k,r}(\{m_j\},\{s_{m}\},\nu_{l},d) 
~I_{k,~r},
\label{NconnectR}
\end{equation}
where $I_{k,r}$  stands for integrals with  $r$ internal lines, 
arbitrary powers of  propagators $\nu_j$ and arbitrary shifts 
of the space-time  dimension $d$ ; $\{s_{q}\}$ is a set of independent
scalar invariants that may be formed from the external momenta.
$Q_i,~R_k$ are ratios of polynomials depending on
$\{s_{r}\}$,  masses $m_j$ , $\nu_{l}$ and  $d$.  
On the left-hand side of (\ref{NconnectR}) we combined 
integrals with  $n$ internal lines and on the right hand side 
integrals with a lesser number of lines.

The key idea in the  derivation of the functional equations is 
to remove integrals with the maximal  number of lines from 
the relations 
(\ref{NconnectR}) by  an appropriate choice of scalar invariants 
$\{s_{q}\}$,  masses $m_j^2$, powers of propagators $\nu_j$ and 
space - time dimension $d$.

In order to obtain functional  equations from a given equation 
(\ref{NconnectR}) one should first solve the polynomial system 
of  equations 
\begin{equation}
 Q_i(\{m_j \},\{s_{q}\},\nu_{l},d)=0,~~~~~~~~i=1,{\ldots} ,N
\end{equation}
with respect to $\{s_{q}\}, \{m_i \},\nu_{l},d$ and then take 
those solutions for which not all coefficients in front of integrals
on the right-hand side of Eq.(\ref{NconnectR}) 
are vanishing. In many cases functional equations can be obtained 
from (\ref{NconnectR})  by choosing only  kinematic  variables 
$\{s_{q}\}$ and masses $\{m_i \}$.

We  illustrate the method by considering the derivation of the 
functional  equations for a one-loop integral  depending on 
$n-1$ independent  external momenta:
\begin{equation}
I_n^{(d)}(\{m_l^2\};~\{p_{ir}\})=
\int \frac{d^d q}{i \pi^{{d}/{2}}} \prod_{j=1}^{n}
\frac{1}{[(q-p_j)^2-m_j^2]^{\nu_j}  },
\end{equation}
where
\begin{equation}
p_{ir}=(p_i-p_r)^2.
\end{equation}
Here and below, the usual causal prescription for the propagators
is understood, i.e. $1/[q^2-m^2] \leftrightarrow 1/[q^2-m^2+i0]$.
Any relation of the form (\ref{NconnectR}) can be used for deriving 
functional equations. In this paper we will use the following relation
\cite{Tarasov:1996br}, \cite{Fleischer:1999hq}:
\begin{eqnarray}
&&G_{n-1} \nu_j{\bf j^+}I^{(d+2)}_n(\{m_l^2\};\{p_{ir}\})
-(\partial_j  \Delta_n)I^{(d)}_n(\{m_l^2\};\{p_{ir}\}) 
\nonumber
\\
&&~~~~~~~~~~~~~~~~~~~~~~~~~~~~~~~~~
=\sum_{k=1}^{n}
 (\partial_j \partial_k \Delta_n)
 {\bf k^-}I^{(d)}_n(\{m_l^2\};\{p_{ir}\}),
\label{reduceJandDtod}
\end{eqnarray}
where the operators   ${\bf j^{\pm }}$ etc. shift the
indices $\nu_j \to \nu_{j } \pm 1$, $G_{n-1}$ is the Gram 
determinant
\begin{equation}
G_{n-1}= -2^n \left|
\begin{array}{cccc}
  (p_1-p_n)(p_1-p_n)  & (p_1-p_n)(p_2-p_n)  
   &\ldots & (p_1-p_n)(p_{n-1}-p_n) \\
  (p_1-p_n)(p_2-p_n)  & (p_2-p_n)(p_2-p_n)  
  &\ldots & (p_2-p_n)(p_{n-1}-p_n) \\
  \vdots  & \vdots  &\ddots & \vdots \\
  (p_1-p_n)(p_{n-1}-p_n)  & (p_2-p_n)(p_{n-1}-p_n)  
           &\ldots & (p_{n-1}-p_n)(p_{n-1}-p_n)
\end{array}
\right|,
\label{Gn}
\end{equation}
and $\Delta_n$ is the modified Cayley determinant defined as:
$$
\Delta_n=  \left|
\begin{array}{cccc}
2m_1^2~~~~  &~~~~m_1^2+m_2^2-p_{12}  &~~\ldots~~ &~~~~ m_1^2+m_n^2-p_{1n} \\
m_1^2+m_2^2-p_{12}~~~   &~~ 2m_2^2  &~~\ldots~~ &~~~~ m_2^2+m_n^2-p_{2n}  \\
\vdots  & \vdots  &~~\ddots & \vdots \\
m_1^2+m_n^2-p_{1n}~~~  & ~~~~m_2^2+m_n^2-p_{2n}&~~\ldots~~   &2m_n^2
\end{array}
         \right|,
$$
\begin{equation}
 \partial_j \equiv \frac{\partial }{ \partial m_j^2}.
\end{equation}
%\begin{equation}
%p_{ij}=(p_i-p_j)^2.
%\label{pij}
%\end{equation}
We assume that the external momenta 
are not restricted  to some specific integer dimension
and therefore $G_{n-1}$ and  $ \Delta_n$ do not 
satisfy any condition specific to a particular value
of the space-time dimension.

In the present paper we will consider functional equations only
for integrals $I_n^{(d)}$ with the first powers of propagators.
Setting all $\nu_k=1$ in Eq.  (\ref{reduceJandDtod}) yields an
equation  of the form (\ref{NconnectR}) connecting integrals
with $n$ and $n-1$ lines.   Functional equations for the integral
$I_{n-1}^{(d)}$ can be obtained for each particular $j$ by 
imposing two conditions:
\begin{equation}
G_{n-1} =0,~~~~~~~~~~~~\partial_j  \Delta_n=0,
\label{sharik}
\end{equation} 
and solving them by an appropriate choice of scalar invariants $p_{ij}$
and masses. There are only $n-1$ independent systems of  relations 
of the type (\ref{sharik}), because
\begin{equation}
\sum_{k=1}^n \partial_k ~\Delta_n = -G_{n-1}.
\end{equation}
Since  $G_{n-1}$  and  $ \partial_j \Delta_n$ are nonlinear in $p_{ij}$ 
and masses, each system of equations may have several solutions. 
The number of functional equations is less than the number 
of possible solutions. This is firstly because  coefficients in front 
of integrals on both sides of Eq.  (\ref{NconnectR}) are simultaneously 
zero for some solutions, and  secondly, because not all functional
 equations are independent.

%%%%%%%%%%%%%%%%%%%%%%%%%%%%%%%%%%%%%%%%%%%%%%%%%%%%%%%%%%%%%%%%%%
\section{Functional equations for the one-loop 
propagator - type integral }
% $I_2^{(d)}(m_1^2,m_2^2;~p_{12})$ }

In accordance with our method described in the previous section,
functional equations for the integral  $I^{(d)}_{2}$
can be obtained from equation (\ref{reduceJandDtod}) 
taken at $n=3$, $~\nu_1=\nu_2=\nu_3=1$.  We will not derive all
possible functional equations, restricting ourselves 
only to the case  $j=1$  in (\ref{reduceJandDtod}):
\begin{eqnarray}
G_2{\bf 1^+}I_3^{(d+2)}(m_1^2,m_2^2,m_3^2;~p_{23},p_{13},p_{12})
&-&(\partial_1\Delta_3) 
I_3^{(d)}(m_1^2,m_2^2,m_3^2;~p_{23},p_{13},p_{12})
\nonumber 
\\
&=&
2(p_{12}+p_{23}-p_{13})I_2^{(d)}(m_1^2,m_2^2;~p_{12})
\nonumber 
\\
&+&2(p_{13}+p_{23}-p_{12})I_2^{(d)}(m_1^2,m_3^2;~p_{13})
\nonumber 
\\
&-&4p_{23}I_2^{(d)}(m_2^2,m_3^2;~p_{23}),
\label{I3intoI2}
\end{eqnarray}
where
\begin{eqnarray}
&&I_3^{(d)}(m_j^2,m_k^2,m_l^2;p_{kl},p_{jl},p_{jk})\!=\!
\int \frac{d^d q}{i \pi^{{d}/{2}}}
\frac{1}{[(q-p_j)^2\!-\!m_j^2]
         [(q-p_k)^2\!-\!m_k^2]
         [(q-p_l)^2\!-\!m_l^2]},
\nonumber
\\
&&I_2^{(d)}(m_j^2,m_k^2;~p_{jk})=
\int \frac{d^d q}{i \pi^{{d}/{2}}}
\frac{1}{[(q-p_j)^2-m_j^2]
         [(q-p_k)^2-m_k^2]}.
\end{eqnarray}
In order to remove integrals $I_3^{(d)},I_3^{(d+2)} $ from this
relation two conditions must be fulfilled:
\begin{eqnarray}
G_{2}&=&2p_{12}^2+2p_{13}^2+2p_{23}^2-4p_{12}p_{13}
-4p_{12}p_{23}-4p_{13}p_{23}=0,
\nonumber
\\
\partial_1 \Delta_3 &=&
  2 p_{23}( p_{13} + p_{12}- p_{23})
 - 4 m_1^2 p_{23} 
\nonumber
\\
&&~~ + 2 m_2^2 (p_{23}+ p_{13}-p_{12})
 +2 m_3^2 (p_{23} +  p_{12} - p_{13})=0.
\label{condition1}
\end{eqnarray}
One can solve this system of equations with respect to
$p_{13}$ and $p_{23}$.
The nontrivial solutions of  the system (\ref{condition1}) 
are:
\begin{eqnarray}
&&p_{13}=s_{13}(m_1^2,m_2^2,m_3^2,p_{12})=
\frac{ \Delta_{12} +2p_{12}(m_1^2+m_3^2)
 - (p_{12}+m_1^2-m_2^2) \lambda}
{2p_{12}},
\nonumber
\\
&&p_{23}=s_{23}(m_1^2,m_2^2,m_3^2, p_{12})=
\frac{ \Delta_{12} +2p_{12}(m_2^2+m_3^2)
 + (p_{12}-m_1^2+m_2^2) \lambda}
{2p_{12}},
\label{solution2}
\end{eqnarray}
where
\begin{equation}
\lambda=\pm \sigma(p_{12}-m_1^2+m_2^2) 
~\sqrt{\Delta_{12}+4p_{12}m_3^2},
\end{equation}
\begin{equation}
\sigma(x)= \left\{ 
\begin{array}{ll}
 +1 & ~~~~\textrm{if} ~~~~x \geq 0  \textrm{,} \\
-1 &  ~~~~\textrm{if} ~~~~x < 0 \textrm{,}
\end{array} \right.
\end{equation}
\begin{equation}
\Delta_{ij}=p_{ij}^2+m_i^4+m_j^4-2p_{ij}m_i^2-2p_{ij}m_j^2
-2m_i^2m_j^2.
\end{equation}
Substituting  Eq.  (\ref{solution2})  into  Eq. (\ref{I3intoI2}) 
yields the following relation:
\begin{eqnarray}
I_2^{(d)}(m_1^2,m_2^2;~p_{12}) =&&
\frac{p_{12}+m_1^2-m_2^2 - \lambda}
 {2p_{12}}
~I_2^{(d)}(m_1^2,m_3^2;~s_{13}(m_1^2,m_2^2,m_3^2,p_{12}))
\nonumber 
\\
 +&&\frac{p_{12}-m_1^2+m_2^2 + \lambda}
 {2p_{12}}
~I_2^{(d)}(m_2^2,m_3^2;~s_{23}(m_1^2,m_2^2,m_3^2,p_{12})).
\label{funcI2short}
\end{eqnarray}
 All arguments of the integral  $I_2^{(d)}$ on the left - hand 
side of  (\ref{funcI2short}) are arbitrary. At the same time, 
the last argument in integrals on the right - hand side  satisfy 
conditions (\ref{condition1}). The mass $m_3$  in the equation 
is an arbitrary parameter and can be chosen at will.
Setting  $m_3=0$ in Eqs. (\ref{solution2})-(\ref{funcI2short}),
 yields
\begin{eqnarray}
\label{mm3zero}
I_2^{(d)}(m_1^2,m_2^2;~p_{12}) &=&
\frac{p_{12}+m^2_1-m^2_2-\alpha_{12}}{2p_{12}}
~I_2^{(d)}(m_1^2,0;~s_{13})
\nonumber
\\
&+&
\frac{p_{12}-m^2_1+m^2_2+\alpha_{12}}{2p_{12}}
~I_2^{(d)}(0,m_2^2;~s_{23}),
\label{equI2_with_mm3z}
\end{eqnarray}
where
\begin{eqnarray}
&&s_{13}=\frac{\Delta_{12}+2p_{12}m_1^2 
-(p_{12}+m_1^2-m_2^2)\alpha_{12}}
{2p_{12}},
\nonumber
\\
&&s_{23}=\frac{ \Delta_{12} +2p_{12} m_2^2
+(p_{12}-m_1^2+m_2^2) \alpha_{12} }
{2p_{12}},
\label{s13_s23}
\end{eqnarray}

\begin{equation}
\alpha_{12}=\pm \sigma(p_{12}-m_1^2+m_2^2)~ \sqrt{\Delta_{12}}~.
\label{alfa}
\end{equation}

The analytic expression for the integral  $I_2^{(d)}(0,m^2;~p^2)$
is  \cite{Bollini:1972bi}, \cite{Boos:1990rg}:
\begin{equation}
I_2^{(d)}(0,m^2;~p^2) = I_2^{(d)}(0,m^2;~0) 
\Fh21\Fup{1,2-\frac{d}{2}}{\frac{d}{2}},
\label{fermprop}
\end{equation}
where
\begin{equation}
I_2^{(d)}(0,m^2;~0)=- \Gamma\left(1-\frac{d}{2}\right) m^{d-4}.
\end{equation}
Thus, relations (\ref{equI2_with_mm3z}) and  (\ref{fermprop}) 
give us the analytic result for the integral $I_2^{(d)}$
with arbitrary masses and external momentum squared.
Our result is in agreement with that presented
in  \cite{Boos:1990rg}.

Setting $m_2^2=0$  in equation (\ref{mm3zero}), assuming 
$|p_{12}|>m_1^2$ and taking solution (\ref{s13_s23}) corresponding 
to the  $+$ sign  in formula (\ref{alfa}) yields
%\begin{equation}
%s_{13}=\frac{m_1^4}{p_{12}},~~~
%s_{23}=\frac{(p_{12}-m_1^2)^2}{p_{12}}.
%\end{equation}
\begin{equation}
I_2^{(d)}(m_1^2,0;~p_{12})=
\frac{m_1^2}{p_{12}}
 I_2^{(d)}\left(m_1^2,0;~ \frac{m_1^4}{ p_{12}} \right)
+\frac{(p_{12}-m_1^2)}{p_{12}}
I_2^{(d)}\left(0,0;~\frac{(p_{12}-m_1^2)^2}{p_{12}}\right).
\label{I2inversearg}
\end{equation}
The first term on the right - hand side  is the same integral  
$I_2^{(d)}$ as on the left - hand side, but with the last argument 
inverted. The second term corresponds to the simple integral  
$I_2^{(d)}$ with both propagators massless:
\begin{equation}
I_2^{(d)}(0,0;~p^2)=
\frac{1}{i \pi^{d/2}} \int \frac{d^dk_1}{k_1^2(k_1-p)^2}=
 \frac{-\pi^{\frac32}~(-p^2)^{\frac{d}{2}-2}}
{2^{d-3}  \Gamma\left(\frac{d-1}{2}\right)
 \sin \frac{\pi d }{2}}.
\label{I2_00p}
\end{equation}
Formula (\ref{I2inversearg}) can be applied to the analytic
continuation of the integral $I_2^{(d)}(m_1^2,0;~p_{12})$ 
into the region of large momenta  $|p_{12}|>m_1^2$. 
It can also be used for the analytic continuation of the integrals 
$I_2^{(d)}$ on  the right - hand side of (\ref{mm3zero}). 
Therefore, the relations (\ref{mm3zero})
and (\ref{I2inversearg}) describe the integral $I_2^{(d)}$
with arbitrary masses and momenta in the whole kinematic
region.

It is interesting to note that equation (\ref{I2inversearg}) 
corresponds to the well-known formula for the analytic
continuation of  Gauss's hypergeometric function
(\ref{fermprop}) (see, for example, Ref.\cite{Erdely}) :
\begin{equation}
\Fh21\Fe{1,2-\frac{d}{2}}{\frac{d}{2}}
=\frac{1}{z} ~\Fh21\Fiz{1,2-\frac{d}{2}}{\frac{d}{2}}
+\frac{\Gamma\left(\frac{d}{2}\right) 
         \Gamma\left(\frac{d}{2}-1\right)}{\Gamma(d-2)}
(-z)^{\frac{d}{2}-2}\left(1-\frac{1}{z}\right)^{d-3}.
\label{F21_largeArg}
\end{equation}
Indeed, substituting the explicit expressions  (\ref{fermprop}),
(\ref{I2_00p}) into (\ref{I2inversearg}) and canceling common factors 
we obtain relation (\ref{F21_largeArg}) with $z=p_{12}/m_1^2$.
%%%%%%%%%%%%%%%%%%%%%%%%%%%%%%%%%%%%%%%%%%%%%%%%%%%%%%%%%%%%%%%%%%

%%%%%%%%%%%%%%%%%%%%%%%%%%%%%%%%%%%%%%%%%%%%%%%%%%%%%%%%%%%%%%%%%%%%%
%%%%%%%%%%%%%%%%%%%%%%%%%%%%%%%%%%%%%%%%%%%%%%%%%%%%%%%%%%%%%%%%%%%%%
\section{Functional equations for the one-loop vertex - type
integral }
%\\ $I_3^{(d)}(m_1^2,m_2^2,m_3^2;~p_{23},p_{13},p_{12})$}
%%%%%%%%%%%%%%%%%%%%%%%%%%%%%%%%%%%%%%%%%%%%%%%%%%%%%%%%%%%%%%%%%%%%%
%%%%%%%%%%%%%%%%%%%%%%%%%%%%%%%%%%%%%%%%%%%%%%%%%%%%%%%%%%%%%%%%%%%%%
Functional equations for the vertex - type integral $I_3^{(d)} $ will
be derived in the same fashion as  for the propagator - type
integral. Setting  $n=4$,  $\nu_1={\ldots} =\nu_4=1$ and $j=1$ 
in Eq.(\ref{reduceJandDtod}) yields:
\begin{eqnarray}
&&G_3~{\bf 1^+}I_4^{(d+2)}(\{m_l^2\};~\{p_{ij}\} )
 - (\partial_1 \Delta_4)I_4^{(d)}(\{m_l^2\};~\{p_{ij}\})= 
\nonumber
\\
&&
~~(\partial_1^2 \Delta_4)
~ I_3^{(d)}(m_2^2,m_3^2,m_4^2;~ p_{34},p_{24},p_{23})
\nonumber 
\\
&+&(\partial_1 \partial_2 \Delta_4)
~I_3^{(d)}(m_1^2,m_3^2,m_4^2; ~p_{34},p_{14},p_{13})
\nonumber
\\
&+&(\partial_1\partial_3 \Delta_4)
~I_3^{(d)}(m_1^2,m_2^2,m_4^2 ;~ p_{24},p_{14},p_{12})
\nonumber
\\
&+&(\partial_1\partial_4 \Delta_4)
~I_3^{(d)}(m_1^2,m_2^2,m_3^2 ;~ p_{23},p_{13},p_{12}).
\label{I4intoI3}
\end{eqnarray}
One can obtain a functional equation for
$I_3(m_1^2,m_2^2,m_3^2;~p_{23},p_{13},p_{12})$ with 
arbitrary arguments by appropriately choosing  
the four variables: $p_{14},p_{24},p_{34}$, $m_4^2$.
To remove the integrals  $I_4^{(d)}, I_4^{(d+2)}$ 
 from  (\ref{I4intoI3}), two conditions
should be satisfied 
\begin{equation}
G_3=0,~~~~\partial_1 \Delta_4=0.
\label{big_sist_4_I3}
\end{equation}
This system of equations depends on 10 variables
 $p_{12},p_{13},p_{14},p_{23},p_{24},p_{34},$
$m_1^2,m_2^2,m_3^2,m_4^2$ and it can be solved 
by excluding, for example,  $p_{14}$ and $p_{34}$.
There are four solutions of the system (\ref{big_sist_4_I3})
but appropriate expressions are rather long and for this 
reason they will not be presented here.
Instead we consider simplified situation, namely we set
in Eq. (\ref{I4intoI3}) from the very beginning   $m_4^2=0$, 
and impose the following conditions
\begin{equation}
G_3=0, ~~~~~~~~~\partial_1 \Delta_4=0,
       ~~~~~~~~~\partial_1\partial_2 \Delta_4=0.
\label{simplified_sist}
\end{equation}
This system can be solved by an appropriate  choice of
$p_{14},p_{34},p_{24}$. There are several solutions of 
(\ref{simplified_sist}), but only for two of them are coefficients in front
of integrals on the right hand side of  (\ref{I4intoI3}) 
different from  zero.
These solutions are
\begin{eqnarray}
\label{solu_mm4zero}
p_{14}&=&s_{14}^{(13)},
\nonumber 
\\
 p_{34}&=&s_{34}^{(13)},
\nonumber
\\
p_{24}&=&s_{24}(m_1^2,m_3^2,p_{23},p_{13},p_{12} )
\nonumber
\\
&=&\frac{ (p_{12}+p_{23}-m_1^2-m_3^2)p_{13}
+(p_{12}-p_{23}-m_1^2+m_3^2)(m_3^2-m_1^2 +\alpha_{13})}{2p_{13}},
\end{eqnarray}
where
\begin{eqnarray}
s_{14}^{(ij)}&=&
 \frac{\Delta_{ij}+2m_i^2 p_{ij}-(p_{ij}+m_i^2-m_j^2)\alpha_{ij} }
 {2p_{ij}},
\nonumber
\\
s_{34}^{(ij)}&=& \frac{\Delta_{ij}+2m_j^2 p_{ij}
+(p_{ij}+m_j^2-m_i^2) \alpha_{ij} }{2p_{ij}},
\nonumber
%\end{eqnarray}
%\begin{eqnarray}
\\
\alpha_{ij}&=&\pm\sigma(p_{ij}-m_i^2+m_j^2) \sqrt{\Delta_{ij}}.
\end{eqnarray}
Substituting  (\ref{solu_mm4zero}) into (\ref{I4intoI3})
leads to the following functional 
equation:
\begin{eqnarray}
 &&I_3^{(d)}(m_1^2,m_2^2,m_3^2;~p_{23},p_{13},p_{12})=
\nonumber
\\
&&~~
\frac{ p_{13}+m_3^2-m_1^2+\alpha_{13} }{2p_{13}}
~ I_3^{(d)}(m_2^2,m_3^2,0;~ 
   s_{34}^{(13)},
   s_{24}(m_1^2,m_3^2,p_{23},p_{13},p_{12}),  p_{23})
\nonumber 
\\
&&~+
\frac{ p_{13}-m_3^2+m_1^2-\alpha_{13} }{2p_{13}}
~I_3^{(d)}(m_1^2,m_2^2,0;~
          s_{24}(m_1^2,m_3^2,p_{23},p_{13},p_{12}),
          s_{14}^{(13)},p_{12}).
\label{mm4zero}
\end{eqnarray}
Relation  (\ref{mm4zero}) means that the integral 
$I_3^{(d)}$ with arbitrary arguments can  always be 
expressed in terms of integrals with at least one massless propagator. 
The only exceptional case, when $p_{12}=p_{13}=p_{23}=0$, is trivial.
In turn, integrals $I_3^{(d)}$ with one massless propagator
can be represented as a sum over integrals with two massless
propagators. Indeed, setting $m_2^2=0$  in Eq. (\ref{mm4zero})  
yields:
\begin{eqnarray}
&&I_3^{(d)}(m_1^2,0,m_3^2;~p_{23},p_{13},p_{12})=
\nonumber
\\
&&~~~~~~ \frac{p_{13}-m_1^2+m_3^2+\alpha_{13}}{2 p_{13}}
 ~I_3^{(d)}(0,m_3^2,0;~s_{34}^{(13)},
           s_{24}(m_1^2,m_3^2,p_{23},p_{13},p_{12}),p_{23})
\nonumber
\\
&&~~~~~~+\frac{p_{13}+m_1^2-m_3^2-\alpha_{13}}{2 p_{13}}
 ~I_3^{(d)}(m_1^2,0,0;~s_{24}(m_1^2,m_3^2,p_{23},p_{13},p_{12}),
      s_{14}^{(13)},p_{12}).
\label{1zer_into_2zer}
\end{eqnarray}
Taking into account the symmetry of the integral $I_3^{(d)}$
with respect to its arguments
one can use Eq. (\ref{1zer_into_2zer}) to express integrals on the 
right hand side of  (\ref{mm4zero}) in terms of integrals with
two propagators massless. Thus in case when
external momenta squared are different from zero the
following relation holds:
\begin{eqnarray}
&&I_3^{(d)}(m_1^2,m_2^2,m_3^2;~p_{23},p_{13},p_{12})=
\nonumber
\\
&& \nonumber \\
&&
  \frac{(p_{13}+m_3^2-m_1^2+\alpha_{13})
         ( p_{23}+m_3^2-m_2^2+\alpha_{23})}
    {4 p_{13} p_{23}}
\nonumber
\\
&&\times
  I_3^{(d)}(m_3^2,0,0;~
   s_{24}(m_2^2,m_3^2,s_{34}^{(13)},p_{23},
   s_{24}(m_1^2,m_3^2,p_{23},p_{13},p_{12})),
   s_{34}^{(23)},
   s_{34}^{(13)})
\nonumber
\\ 
&& \nonumber \\
&& \nonumber \\
&&
  +\frac{(p_{13}+m_3^2-m_1^2+\alpha_{13})
         (p_{23}-m_3^2+m_2^2-\alpha_{23})}
    {4p_{13}p_{23}}
\nonumber
\\
&&\times I_3^{(d)}(m_2^2,0,0;~
  s_{24}(m_2^2,m_3^2,s_{34}^{(13)},p_{23},
  s_{24}(m_1^2,m_3^2,p_{23},p_{13},p_{12})),
  s_{14}^{(23)},
  s_{24}(m_1^2,m_3^2,p_{23},p_{13},p_{12}))
\nonumber
\\  
&& \nonumber \\
&&  +\frac{(p_{13}-m_3^2+m_1^2-\alpha_{13})
          (p_{12}+m_2^2-m_1^2+\alpha_{12})}
     {4p_{13} p_{12}}
\nonumber
\\
&&\times
 I_3^{(d)}(m_2^2,0,0;~
  s_{24}(m_1^2,m_2^2,s_{24}(m_1^2,m_3^2,p_{23},p_{13},p_{12}),p_{12},
  s_{14}^{(13)}),
  s_{34}^{(12)},
  s_{24}(m_1^2,m_3^2,p_{23},p_{13},p_{12}))
\nonumber
\\
&& \nonumber \\
&&  
  +\frac{(p_{13}-m_3^2+m_1^2-\alpha_{13})
 (p_{12}-m_2^2+m_1^2-\alpha_{12})}{4 p_{12} p_{13}}
\nonumber
\\
&&\times
  I_3^{(d)}(m_1^2,0,0;~
  s_{24}(m_1^2,m_2^2,s_{24}(m_1^2,m_3^2,p_{23},p_{13},p_{12}),
     p_{12},s_{14}^{(13)}),
     s_{14}^{(12)},
     s_{14}^{(13)}).
\end{eqnarray}  
There is one further simplification of note.
Setting  $m_1^2=m_3^2=p_{24}=0$ in Eq. (\ref{I4intoI3}) from the very   
beginning  and solving  system of equations
\begin{equation}
G_3=0,~~~~~~\partial_1 \Delta_4=0,
\end{equation}
with respect to $p_{14}$ and $p_{34}$ yields a nontrivial relationship:
\begin{eqnarray}
&&I_3^{(d)}(0,m^2,0;~p_{23},p_{13},p_{12})=
\nonumber
\\
&&-\frac{[b(p_{23},p_{12})-(p_{12}+m^2) \alpha_{123}]m^2}
    {2 \Lambda_3}~
I_3^{(d)}(m^2,0,0;~\kappa_{14},0,p_{12})
\nonumber
\\
&&
 -\frac{[b(p_{12},p_{23})+(p_{23}+m^2) \alpha_{123}]m^2}{2\Lambda_3}
 ~ I_3^{(d)}(m^2,0,0;~\kappa_{34},0,p_{23})
\nonumber 
\\
&& 
+ \frac{\Lambda_3 - p_{13} (p_{12} p_{23}-m^4)
  - m^2 (p_{12}-p_{23}) \alpha_{123} }{2 \Lambda_3}
~I_3^{(d)}(0,0,0;~\kappa_{34},\kappa_{14},p_{13}),
\label{I3zmzppp}
\end{eqnarray} 
where
\begin{eqnarray}
&&\kappa_{14} = 
\frac{a(p_{23},p_{12})+m^2 b(p_{23},p_{12})\alpha_{123}}
  {2 \Lambda_3},
\nonumber
\\
&&\nonumber \\
&&\kappa_{34} = 
\frac{a(p_{12},p_{23})-m^2 b(p_{12},p_{23}) \alpha_{123}}
  {2 \Lambda_3},
\end{eqnarray}
\begin{eqnarray}
&&\Lambda_3=(m^4-p_{12}p_{23})(p_{23}-p_{12})-(p_{12}+m^2)
b(p_{12},p_{23}), 
\nonumber
\\
&&\nonumber \\
&&
a(p_{12},p_{23})=
m^2(p_{12}-p_{23})^2 (p_{23}-m^2)
+[2 (m^2-p_{23}) p_{12}-m^2 p_{13}] (p_{23}+m^2)p_{13},
\nonumber \\
&&\nonumber \\
&&b(p_{12},p_{23})=
(m^2+p_{13}+p_{12}-p_{23})p_{23} +m^2 (p_{13}-p_{12}),
\nonumber
\\
&&\nonumber \\
&&\alpha_{123}=\sigma(b(p_{12},p_{23})) \sqrt{\Delta_{123}},
\nonumber 
\\
&& \nonumber \\
&& \Delta_{123}= p_{12}^2+p_{13}^2+p_{23}^2
-2p_{12}p_{13}-2p_{12}p_{23}-2p_{13}p_{23}.
\end{eqnarray}
Therefore, by using Eq.(\ref{I3zmzppp}) we can represent
the integral $I_3^{(d)}$ with arbitrary masses and nonzero 
kinematic variables as a combination of integrals with
two massless propagators, one momentum squared equal to zero
and integrals with all propagators massless.
An analytic result for the integral $I_3^{(d)}$ with all 
propagators massless  is known in terms  of $_2F_1$ functions
(see Ref. \cite{Davydychev:1999mq}).

Integrals $I_3^{(d)}$ with two massless propagators on the right - hand
side of (\ref{I3zmzppp}) can be evaluated analytically. In the kinematic
region  $|p_{12}|\leq m^2$ and $|p_{13}|\leq m^2$ we find
\begin{eqnarray}
&&I_3^{(d)}(0,m^2,0;~0,p_{13},p_{12})=
-\frac{I_2^{(d)}(0,0;~p_{13}) }
{m^2}\Fh21\FpR{1,\frac{d-2}{2}}{d-2}
\nonumber
\\
&&~~~~~+\frac{1}{m^2}~I_2^{(d)}(0,m^2;~0)
~F_1\left(1,1,2-\frac{d}{2}, \frac{d}{2};
  \frac{p_{12}-p_{13}}{m^2}, \frac{p_{12}}{m^2}\right),
\label{AppellF1}
\end{eqnarray}
where $F_1$ is the Appell hypergeometric function \cite{ApKdF}  
which  admits a simple  one-fold integral representation:
\begin{equation}
F_1\left(1,1,2-\frac{d}{2},\frac{d}{2};x,y\right)=
\frac{(d-2)}{2}\int_0^1du\frac{[(1-u)(1-yu)]^{\frac{d}{2}-2}}
{(1-xu)} .
\end{equation}
Thus by using  (\ref{AppellF1}) one can obtain the result 
for the integral  $I_3^{(d)}$ in terms of the Appell function $F_1$ 
and Gauss's hypergeometric function $_2F_1$.
This result is in agreement with the result obtained in Ref.
\cite{Fleischer:2003rm} and later in Ref. \cite{Davydychev:2005nf}. 
At  $d=4$ the result for  $I_3^{(4)}$ in terms of Appell function
$F_3$ was obtained in Ref. \cite{CabralRosetti:1998sp}.

The Appell function  $F_1$ in formula (\ref{AppellF1})
has branch points if  
\begin{equation}
|p_{12} | \geq m^2,~~~~~~~~~~
{\rm or}~~~~~~~~~~ |p_{12}-p_{13}| \geq m^2.
\label{LargeArgs}
\end{equation}
To analytically continue the integral 
$I_3^{(d)}(0,m^2,0;~0,p_{13},p_{12})$ into regions 
(\ref{LargeArgs}) one can use appropriate functional 
equations. When  $|p_{12}| \geq m^2$,   the following
functional equation can be applied: 
\begin{eqnarray}
&&
I_3^{(d)}(0,m^2,0;~0,p_{13},p_{12})=
 \frac{m^2}{p_{12}}~
I_3^{(d)}\left(0,m^2,0;~0,
\frac{m^2(p_{13}-p_{12}+m^2)}{p_{12}},\frac{m^4}{p_{12}}\right)
\nonumber
\\
&&~~~~~~~
+\frac{(p_{12}-m^2)}{p_{12}}~
I_3^{(d)}\left(0,0,0;~\frac{m^2(p_{13}-p_{12}+m^2)}{p_{12}},
\frac{ (p_{12}-m^2)^2}{p_{12}},p_{13}\right).
\label{p12continued}
\end{eqnarray}
This relation can be derived  from Eq. (\ref{I4intoI3}) with
$m_1^2=m_3^2=m_4^2=p_{23}=0$ and $m_2^2=m^2$   by imposing 
the following conditions:
\begin{equation}
G_4=0,~~~~\partial_1\Delta_4=0,~~~~\partial_1\partial_3 \Delta_4=0.
\end{equation}
On the right - hand side of the relation (\ref{p12continued})
the last two arguments of the integral $I_3^{(d)}$ in the first
term  are finite for large $|p_{12}|$.

If  $|p_{12}-p_{13}| \geq m^2$ and  $|p_{12}| \leq m^2$, 
the following relation can be applied
\begin{eqnarray}
&&I_3^{(d)}(0,m^2,0;~0,p_{13},p_{12})=
\frac{p_{12} m^2}{m^2 p_{13}+p_{12} p_{13}-p_{12}^2}
I_3^{(d)}\left(0,m^2,0;~0,\frac{p_{12}^2(p_{13}-p_{12}+m^2)}
{m^2 p_{13}+p_{12} p_{13}-p_{12}^2},p_{12}\right)
\nonumber
\\
&&
~~~~~~~~
+\frac{p_{12}(p_{13}-p_{12})}{m^2 p_{13}+p_{12} p_{13}-p_{12}^2}~
I_3^{(d)}\left(0,0,0;~\frac{m^2 (p_{13}-p_{12})^2}
{m^2 p_{13}+p_{12} p_{13}-p_{12}^2},\frac{p_{12}^2(p_{13}-p_{12}+m^2)}
{m^2 p_{13}+p_{12} p_{13}-p_{12}^2},p_{13}\right)
\nonumber
\\
&&
~~~~~~~~+
\frac{m^2(p_{13}-p_{12})}{m^2 p_{13}+p_{12} p_{13}-p_{12}^2}
I_3^{(d)}\left(0,m^2,0;~0,\frac{m^2 (p_{13}-p_{12})^2}
{m^2 p_{13}+p_{12} p_{13}-p_{12}^2},0\right).
\label{p13continued}
\end{eqnarray}
This relation can be derived  from Eq. (\ref{I4intoI3}) with
$m_1^2=m_3^2=m_4^2=p_{23}=p_{24}=0$ and $m_2^2=m^2$   
by imposing  the following conditions:
\begin{equation}
G_4=0,~~~~\partial_1\Delta_4=0.
\end{equation}
The penultimate  argument of the first integral on the
right hand side of (\ref{p13continued}) is finite
for large values of $|p_{13}| \geq m^2$ if $p_{12}\neq -m^2$.
The second and the third integrals on the right - hand side 
of this relation can be expressed in terms of the
hypergeometric function $_2F_1$ and their analytic continuation
causes no problems.  

If both conditions (\ref{LargeArgs}) hold then the analytic 
continuation can be done by  applying  both (\ref{p12continued}) 
and (\ref{p13continued}). 
%%%%%%%%%%%%%%%%%%%%%%%%%%%%%%%%%%%%%%%%%%%%%%%%%%%%%%%%%%%%%%%%

\section{Conclusions}

Finally, we summarize what we have accomplished in this paper.

First of all, we formulated the general method for 
deriving functional equations for Feynman integrals.

Second, it was shown that integrals with many
kinematic arguments can be reduced to a combination
of integrals with simpler kinematics.

Third, we demonstrated that our functional equations
can be used for the analytic continuation of Feynman
integrals to  all  kinematic domains.
 
In the present paper we considered rather particular cases 
of functional equations. 
 The systematic investigation and classification of
the proposed functional equations requires application
of the methods of algebraic geometry and group theory.

A detailed consideration of our functional 
equations and their application to the one-loop
integrals with four,  five and six external legs
as well as  to some two- and three- loop  Feynman 
integrals will  be presented in future publications.

\section{Acknowledgment}

I am very thankful to Ronald Reid-Edwards for carefully 
reading the manuscript and useful remarks.
This work was supported  in part from DFG grants  DFG KN365/3 
and  BMBF 05HT6GUA.
Part of this investigation was done during my stay at 
the Institut f\"ur  Theoretische Physik E, RWTH Aachen
where I was supported from the DFG grant 
Sonderforschungsbereich Transregio 9-03.


\begin{thebibliography}{99}

\bibitem{Petersson}
B.~Petersson, {\em J.~Math.~Phys.} {\bf 6},  (1965) 1955.
%\cite{'t Hooft:1972fi}
\bibitem{'t Hooft:1972fi}
  G.~'t Hooft and M.~J.~G.~Veltman,
  %``Regularization And Renormalization Of Gauge Fields,''
  Nucl.\ Phys.\  B {\bf 44},  (1972) 189.
  %%CITATION = NUPHA,B44,189;%%

%\cite{Tkachov:1981wb}
\bibitem{Tkachov:1981wb}
  F.~V.~Tkachov,
  %``A Theorem On Analytical Calculability Of Four Loop Renormalization Group
  %Functions,''
  Phys.\ Lett.\  B {\bf 100} (1981) 65;\\
  %%CITATION = PHLTA,B100,65;%%
%\cite{Chetyrkin:1981qh}
%\bibitem{Chetyrkin:1981qh}
  K.~G.~Chetyrkin and F.~V.~Tkachov,
  %``Integration By Parts: The Algorithm To Calculate Beta Functions In 4
  %Loops,''
  Nucl.\ Phys.\  B {\bf 192} (1981) 159.
  %%CITATION = NUPHA,B192,159;%%

%\cite{Tarasov:1996br}
\bibitem{Tarasov:1996br}
  O.~V.~Tarasov,
  %``Connection between Feynman integrals having different values of the
  %space-time dimension,''
  Phys.\ Rev.\  D {\bf 54} (1996) 6479
  [arXiv:hep-th/9606018].
  %%CITATION = PHRVA,D54,6479;%%


%\cite{Fleischer:1999hq}
\bibitem{Fleischer:1999hq}
  J.~Fleischer, F.~Jegerlehner and O.~V.~Tarasov,
  %``Algebraic reduction of one-loop Feynman graph amplitudes,''
  Nucl.\ Phys.\  B {\bf 566} (2000) 423
  [arXiv:hep-ph/9907327].
  %%CITATION = NUPHA,B566,423;%%


%\cite{Bollini:1972bi}
\bibitem{Bollini:1972bi}
  C.~G.~Bollini and J.~J.~Giambiagi,
  %``Lowest order divergent graphs in nu-dimensional space,''
  Phys.\ Lett.\  B {\bf 40} (1972) 566.
  %%CITATION = PHLTA,B40,566;%%

%\cite{Boos:1990rg}
\bibitem{Boos:1990rg}
  E.~E.~Boos and A.~I.~Davydychev,
  %``A Method of evaluating massive Feynman integrals,''
  Theor.\ Math.\ Phys.\  {\bf 89},  (1991) 1052;
  Teor.\ Mat.\ Fiz.\  {\bf 89},  (1991) 56 .
  %%CITATION = TMFZA,89,56;%%

\bibitem{Erdely} A.~Erd\`ely et. al.,
   {\it Higher Transcendental Functions} 
   Vol.1, McGraw-Hill, New York, 1953. 

%\cite{Davydychev:1999mq}
\bibitem{Davydychev:1999mq}
  A.~I.~Davydychev,
  %``Explicit results for all orders of the epsilon-expansion of certain
  %massive and massless diagrams,''
  Phys.\ Rev.\  D {\bf 61} (2000) 087701
  [arXiv:hep-ph/9910224].
  %%CITATION = PHRVA,D61,087701;%%


\bibitem{ApKdF} P.~Appell and  J.~Kamp\'e de F\'eriet  {\it
Fonctions hypergeometriques et hypersp\'eriques}, 
 Gauthier Villars, Paris,  1926.


%\cite{Fleischer:2003rm}
\bibitem{Fleischer:2003rm}
  J.~Fleischer, F.~Jegerlehner and O.~V.~Tarasov,
  %``A new hypergeometric representation of one-loop scalar integrals in d
  %dimensions,''
  Nucl.\ Phys.\  B {\bf 672},  (2003) 303,
  [arXiv:hep-ph/0307113].
  %%CITATION = NUPHA,B672,303;%%


%\cite{Davydychev:2005nf}
\bibitem{Davydychev:2005nf}
  A.~I.~Davydychev,
  %``Geometrical methods in loop calculations and the three-point function,''
  Nucl.\ Instrum.\ Meth.\  A {\bf 559} (2006) 293
  [arXiv:hep-th/0509233].
  %%CITATION = NUIMA,A559,293;%%

%\cite{CabralRosetti:1998sp}
\bibitem{CabralRosetti:1998sp}
  L.~G.~Cabral-Rosetti and M.~A.~Sanchis-Lozano,
  %``Generalized hypergeometric functions and the evaluation of scalar  one-loop
  %integrals in Feynman diagrams,''
  J.\ Comput.\ Appl.\ Math.\  {\bf 115} (2000) 93
  [arXiv:hep-ph/9809213].
  %%CITATION = JCAMD,115,93;%%




\end{thebibliography}
\end{document}